\documentclass[numberedappendix]{emulateapj}
\usepackage{apjfonts}
\usepackage{graphicx}
\usepackage{psfig}
\usepackage{mathrsfs}

\shorttitle{Hot Galactic Winds and X-ray Emission}
\shortauthors{Zhang et al.}

\begin{document}

%% LaTeX will automatically break titles if they run longer than
%% one line. However, you may use \\ to force a line break if
%% you desire.

\title{Hot Galactic Winds Constrained by the X-Ray Luminosities of Galaxies}
%% Use \author, \affil, and the \and command to format
%% author and affiliation information.
%% Note that \email has replaced the old \authoremail command
%% from AASTeX v4.0. You can use \email to mark an email address
%% anywhere in the paper, not just in the front matter.
%% As in the title, use \\ to force line breaks.

\author{Dong Zhang\altaffilmark{1,2}, Todd A.~Thompson\altaffilmark{1,2}, Norman Murray\altaffilmark{3,5}, and Eliot Quataert\altaffilmark{4}}
\altaffiltext{1}{Department of Astronomy, The Ohio State University,
140 W. 18th Ave., Columbus, OH, 43210, USA; dzhang,
 thompson@astronomy.ohio-state.edu}

\altaffiltext{2}{Center for Cosmology \& Astro-Particle Physics, The
Ohio State University, 191 West Woodruff Ave., Columbus, OH, 43210, USA}

\altaffiltext{3}{Canadian Institute for Theoretical Astrophysics,
60 St. George Street, University of Toronto, Toronto, ON M5S 3H8, Canada}

\altaffiltext{4}{Astronomy Department \& Theoretical Astrophysics Center,
601 Campbell Hall, University of California, Berkeley, CA 94720, USA}

\altaffiltext{5}{Canada Research Chair in Astrophysics}

\begin{abstract}
Galactic superwinds may be driven by very hot outflows generated by overlapping supernovae within the host galaxy. We use the Chevalier \& Clegg (CC85) wind model and the observed correlation between X-ray luminosities of galaxies and their SFRs to constrain the mass-loss rates ($\dot{M}_{\rm hot}$) across a wide range of star formation rates (SFRs), from dwarf starbursts to ultra-luminous infrared galaxies. We show that for fixed thermalization and mass-loading efficiencies, the X-ray luminosity of the hot wind scales as $L_{X}\propto {\rm SFR}^2$, significantly steeper than is observed for star-forming galaxies: $L_{X}\propto {\rm SFR}$.  Using this difference we constrain the mass-loading and thermalization efficiency of hot galactic winds. For reasonable values of the thermalization efficiency ($\lesssim1$) and for ${\rm SFR}\gtrsim10$\,M$_\odot$ yr$^{-1}$ we find that $\dot{M}_{\rm hot}/{\rm SFR}\lesssim1$, significantly lower than required by integrated constraints on the efficiency of stellar feedback in galaxies, and potentially too low to explain observations of winds from rapidly star-forming galaxies.  In addition, we highlight the fact that heavily mass-loaded winds cannot be described by the adiabatic CC85 model because they become strongly radiative.
\end{abstract}


\section{Introduction}\label{Intro}

Galactic-scale winds are important in rapidly star-forming galaxies. They are the primary mechanism by which energy and metals are ejected from galaxies and deposited into the intergalactic medium, and they are a product of the feedback mechanisms at work in regulating star formation.

A number of mechanisms have been proposed for launching galactic superwinds, including energy and momentum deposition by supernovae (SN), radiation pressure on dust, and cosmic rays \citep{dekel_silk, MQT05, Everett08, Socrates08, hopkins_wind}.
%galactic-scale winds may be driven by an initially very hot flow, which is created from the combination of many supernova (SN) explosions in the galaxy and which sweeps up cool clouds via ram pressure acceleration
%\citep{SS00}.
\cite{CC85} (hereafter CC85) developed a one-dimensional (1D) spherically-symmetric model for a very hot wind created by supernova energy injection with two controlling parameters: the thermalization efficiency with which SN energy is converted into thermal energy, and the mass-loading efficiency. Numerical simulations show that this analytic model provides a good approximation in describing the hot wind fluid properties and emission from axisymmetric disk-like configurations (\citealt{SH09}), or in three-dimensional starburst models (\citealt{Stevens03}).

The thermalization efficiency and, in particular, the mass-loading efficiency are the crucial parameters that determine the overall importance of hot winds in driving matter and metals out of galaxies.  Higher thermalization efficiencies imply higher velocities and higher temperatures, and larger mass-loading rates imply more hot wind momentum available to accelerate cold clouds.  Despite their importance these parameters are difficult to determine observationally. Both low (e.g., \citealt{Bradamante98}) and high thermalization efficiency (e.g., \citealt{SS00}) have been inferred. Constraints on the mass-loss rate of the hot flow ($\dot{M}_{\rm hot}$) in individual galaxies by observation are few, e.g., NGC 1569 (\citealt{Martin02}), and M82 (\citealt{SH09}).

One method to constrain the hot wind properties directly is by X-ray observations. Recently, \cite{SH09} constrained the wind parameters in the archetypal nearby starburst galaxy M82 using hard X-ray observations of its central region, finding a high thermalization efficiency ($\sim1$) and a mass-loading efficiency of $\dot{M}_{\rm hot}/{\rm SFR}\sim0.5$. However, superwinds in other galaxies with star formation rates (SFRs) of $1-1000\,M_{\odot}$ yr$^{-1}$ at both low and high redshift are much less well studied, and a more generic approach needs to be introduced to constrain the hot wind properties and to understand their dynamical importance for rapidly star-forming galaxies. Therefore, we apply the CC85 model across a wide range of galaxies from dwarf starbursts to ultra-luminous infrared galaxies (ULIRGs). By using the observed X-ray properties of galaxies we constrain the thermalization efficiency and mass loading of hot winds.

Star-forming galaxies are luminous X-ray emitters. In particular, the X-ray luminosities of star-forming galaxies exhibit a tight linear correlation with their SFRs over about four orders of magnitude from $\sim0.1\,M_{\odot}$ yr$^{-1}$ to $\sim\,10^{3}M_{\odot}$ yr$^{-1}$ (\citealt{Grimm03}; \citealt{Ranalli03}; \citealt{Gilfanov04}; \citealt{Persic07}; \citealt{Dijkstra12}; \citealt{Mineo11, Mineo12a, Mineo12b, Mineo14}). Observationally, X-ray binaries (XRBs) are the benchmark X-ray tracers that give rise to this linear correlation, but the total X-ray luminosity $L_{X}^{\rm tot}$ also has contributions from young SN remnants, neutron stars, the warm/hot ISM, and potentially the very hot coronal gas of a CC85-like galactic outflow. The most recent normalization of the linear correlation between the \textit{total} $L_{X}^{\rm tot}$ and the SFR is given by \cite{Mineo14}, who find that\footnote{Earlier studies yielded slightly different values of the scalefactor between compact X-ray sources ($2-10$ keV) and SFR, including (in unit of erg s$^{-1}$/$(M_{\odot}$ yr$^{-1})$) $0.91\times10^{39}$ (\citealt{Ranalli03}), $1.2\times10^{39}$ (\citealt{Grimm03}), $2.2\times10^{39}$ (\citealt{Shtykovskiy05}), $0.75\times10^{39}$ (\citealt{Persic07}) and $1.7\times10^{39}$ (\citealt{Lehmer10}).}
\begin{equation}
L_{X\,(0.5-8\,\rm{keV})}^{\rm tot}/\textrm{SFR}\simeq(4.0\pm0.4)\times10^{39}\;\textrm{erg}\,\textrm{s}^{-1}/(M_{\odot}\,\textrm{yr}^{-1}),
\label{lxhard}
\end{equation}
where the X-ray emission is from 0.5 keV to 8 keV. Among the multiple contributors to the X-ray emission of galaxies, it is the diffuse emission from hot gas that is of special interest here. As we show in Section \ref{Model}, considering only X-rays produced by thermal emission, the CC85 model predicts that the diffuse hard X-ray emission from hot wind gas should scale as $L_{X}\propto {\rm SFR}^{2}$ at fixed thermalization and mass-loading efficiency. Because the observed correlation is $L_{X}\propto {\rm SFR}$ this provides an observational limit on the contribution of hot diffuse X-ray emission, which can then be used to constrain the wind model.

In Section \ref{Model} we introduce the CC85 model and calculate the hard X-ray emission from the hot wind fluid. In Section \ref{sect_LxSFR} the observed $L_{X}-$SFR relation is used to constrain the thermalization and mass-loading efficiency of hot winds. In Section \ref{discussion} we discuss whether our results change with different parameters in the model and different forms of the observed $L_{X}-$SFR correlation. Conclusions are presented in Section \ref{conclusions}.

%------------------------------------------------------------------------%
\begin{figure}[t]
\centerline{\includegraphics[width=9.3cm]{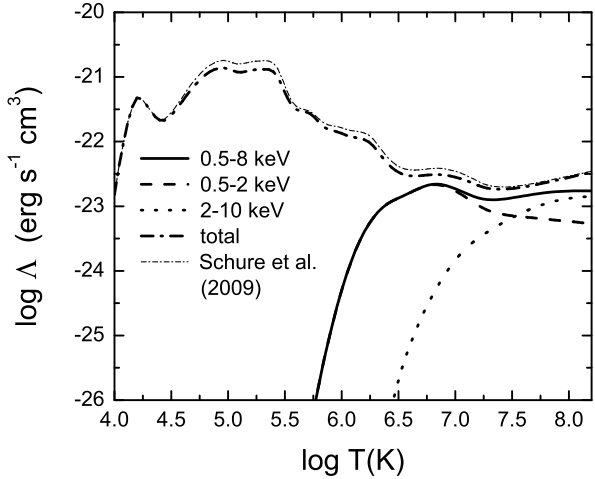}}
\caption{Broadband cooling functions calculated by SPEX in different frequency ranges for solar metallicity. The total cooling curve is calculated by integrating the emissivity from 0.1 eV to 1 MeV.}\label{fig_Cooling}
\end{figure}
%------------------------------------------------------------------------%
%------------------------------------------------------------------------%
%\begin{figure*}[p]
\begin{figure*}[t]
\centerline{\includegraphics[width=13cm]{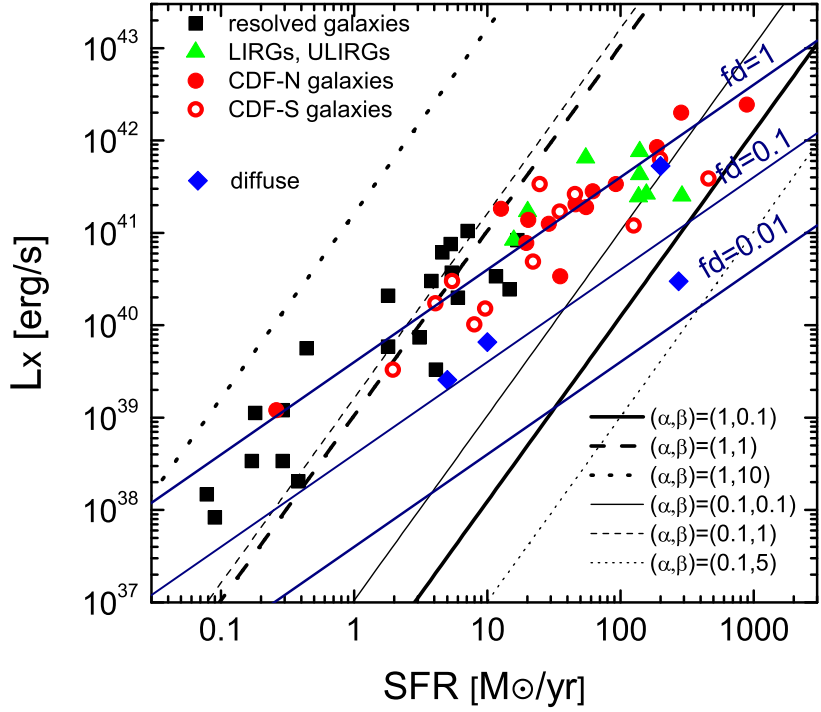}}
\caption{X-ray luminosity $L_{X}$ of the hot wind at 0.5-8 keV ($L_{X,\rm hot}$) with the set of parameters $(\alpha,\beta)=(1,0.1), (1,1), (1,10)$ (thick lines) and $(0.1,0.1), (0.1,1), (0.1,10)$ (thin lines) and assuming $R=200$ pc, compared with the diffuse X-ray luminosity ($L_{X,\rm diffuse}$) from galaxies where the fraction of diffuse X-ray emission $f_{d}=0.01,0.1$ or 1 (blue lines). Black squares, green triangles, and red circles are the total X-ray luminosities of resolved galaxies, LIRGs, ULIRGs, {\em Chandra Deep Field North} (CDF-N) and {\em Chandra Deep Field South} (CDF-S) galaxies in \cite{Mineo14}. The four blue diamonds are the diffuse X-rays (not total) from M82, NGC 253, NGC 6240 and Arp 220 (see Section \ref{discussion}). }\label{fig_Lx}
\end{figure*}
%------------------------------------------------------------------------%

\section{Supernova-Driven Hot Wind Model and Expected X-ray Emission}\label{Model}

\subsection{The CC85 Galactic Wind Model}\label{sectCC85model}

The analytic spherically-symmetric hot flow solution derived by CC85 depends on three parameters: the energy input rate $\dot{E}_{\rm hot}$, mass loss rate $\dot{M}_{\rm hot}$, and the outflow launch radius $R$ which can be de-dimensionalized either by M82's parameters (e.g, CC85 model) or the SFR of galaxies (e.g., \citealt{SH09}). For simplicity, two dimensionless parameters $\alpha'$ and $\beta$ are introduced to normalize the energy input and mass-loading efficiency by
\begin{eqnarray}
\dot{E}_{\rm hot}&=&\alpha'\,\dot{E}_{\rm SN},\label{parameter1}\\
\dot{M}_{\rm hot}&=&\beta\,\textrm{SFR}.\label{parameter2}
\end{eqnarray}
The cumulative net energy input from SNe $(\dot{E}_{\rm SN})$ is given by
\begin{equation}
\dot{E}_{\rm SN}=\epsilon\nu\;\textrm{SFR},
\end{equation}
where $\epsilon=10^{51}\epsilon_{51}$ ergs is the energy injected by an individual SN, SFR$_{0}=$SFR$/(M_{\odot}$ yr$^{-1})$, and $\nu=(100M_{\odot})^{-1}\nu_{100}$ is the number of SNe per unit mass of star formation, where typically one SN occurs per $100M_{\odot}$ of stars produced, i.e., $\nu_{100}\simeq1$. For a Salpeter or Chabrier stellar initial mass function (IMF), $\nu_{100}=1.18$ and $\nu_{100}=1.74$, respectively (\citealt{Leitherer99}, \citealt{SH09}). Thus we use $\alpha=\alpha' \epsilon_{51}\nu_{100}$ to parameterize the energy injection rate. Here $\alpha$ could in principle be as high as a few, depending on the IMF model, and the contribution to the total heating by stellar winds. We take $\alpha\lesssim2$ in this paper.
% Since $\alpha'\leq1$ and $\nu_{100}\sim1$, the plausible range of $\alpha$ covers from $\alpha<1$ to $\alpha\sim1-10$.
Equation (\ref{parameter1}) is then rewritten as
\begin{equation}
\dot{E}_{\rm hot}\simeq 3.2\times10^{41}\,\textrm{erg}\,\textrm{s}^{-1}\;\alpha\,\textrm{SFR}_{0}.\label{parameter3}
\end{equation}

The solution to the CC85 model is given in Appendix \ref{CC85appendix}. In particular, the temperature $T$, density $n$ and velocity $V_{\rm hot}$ of the hot wind outflow are
\begin{eqnarray}
T(r)&=&6.1\times10^{7}\;\textrm{K}\;\mu\left(\frac{\alpha}{\beta}\right)\left[\frac{P_{*}(r_{*})}{\rho_{*}(r_{*})}\right]\label{WindT}\\
n(r)&=&1.4\;\textrm{cm}^{-3}\;\alpha^{-1/2}\beta^{3/2}\mu^{-1}R_{200\rm pc}^{-2}\rho_{*}(r_{*})\textrm{SFR}_{0}\label{winddensity}\\
V_{\rm hot}(r)&=&710\;\textrm{km}\;\textrm{s}^{-1}\;\alpha^{1/2}\beta^{-1/2}u_{*}(r_{*}),\label{Vhot}
\end{eqnarray}
respectively, where $R_{200\rm pc}=R/(200\;\textrm{pc})$, $r$ is the radius of the wind, $r_{*}=r/R$ is the dimensionless radius, $u_{*}$, $\rho_{*}$ and $P_{*}$ are the dimensionless velocity, density as functions of dimensionless radius $r_{*}$ in the CC85 model respectively, and $\mu$ is the mean molecular weight. Note that equations (\ref{AppenP}) and (\ref{Appenrho}) in Appendix \ref{CC85appendix} show that the density of the hot flow is $n\propto\rho\propto\dot{M}_{\rm hot}^{3/2}\dot{E}_{\rm hot}^{-1/2}$. Given fixed $\alpha$ and $\beta$, equations (\ref{parameter1}) and (\ref{parameter2}) then give $\dot{E}_{\rm hot}\propto$ SFR and $\dot{M}_{\rm hot}\propto$ SFR, thus we have $n\propto$ SFR. For solar abundance $\mu\approx0.61$, we have $T=1.5\times10^{7}$ K $(\alpha/\beta)$, $n=1.1$ cm$^{-3}$ $\alpha^{-1/2}\beta^{3/2}R_{200\rm pc}^{-2}\rho_{*(r=0)}\textrm{SFR}_{0}$ at the center of the host galaxy ($r=0$), while $V_{\rm hot}=10^{3}$ km s$^{-1}$ $\alpha^{1/2} \beta^{-1/2}$ at infinity.

\subsection{X-ray Cooling and Emission}

The diffuse X-ray continuum emission from the hot wind fluid is
\begin{equation}
L_{\rm X,hot}^{[\nu_{1},\nu_{2}]}=\int n_{e}n_{\rm H}\Lambda_{\rm N}^{[\nu_{1},\nu_{2}]}(T,Z)dV
\end{equation}
where $\Lambda_{\rm N}^{[\nu_{1},\nu_{2}]}(T,Z)$ is the emissivity at temperature $T$, metallicity $Z$, and in the energy band between two (X-ray) frequencies $\nu_{1}$ and $\nu_{2}$, $n_{\rm H}$ and $n_{e}$ are the hydrogen and electron number density respectively. Standard free-free bremsstrahlung emission dominates the cooling rate $\Lambda_{\rm N}(T,Z)$ for $T\geq10^{7}$ K ($\sim1$ keV). A better cooling model should include line emission from ions. Following \cite{Schure09}, we use the SPEX package\footnote{http://www.sron.nl/spex} (version 2.03.03) to calculate $\Lambda_{\rm N}(T,Z)$ of a hot plasma in collisional ionization equilibrium (CIE). CIE is a valid assumption if the plasma is dominated by collisional processes and the cooling timescale is longer than the recombination or ionization timescale, otherwise non-equilibrium ionization (NEI) treatments should be taken into account. \cite{SH09} found that it is justified to use CIE treatment in M82, and \cite{Schure09} discussed that the differences between NEI and CIE emission are quite small for $T\geq10^{6}$ K. Therefore we adopt CIE treatment. Figure \ref{fig_Cooling} shows the $0.5-8$ keV, $0.5-2$ keV, $2-10$ keV, and the total cooling functions by integrating over the energy range from 0.1 eV to 1 MeV. Solar abundances from \cite{Lodders09} are assumed in this calculation, thus the total cooling function is slightly different from \cite{Schure09} who adopted solar abundances from \cite{Anders89}. The total cooling curve has an order of $ \Lambda_{\rm N}\sim 10^{-22.5}$ erg s$^{-1}$ cm$^{3}$ for $T\geq10^{6}$ K, while the cooling functions at $0.5-8$ keV and $0.5-2$ keV reach maximum $\simeq 10^{-22.7}$ erg s$^{-1}$ cm$^{3}$ at $T\simeq 10^{6.8}$ K. 

The densities $n_{\rm H}$ can be calculated using CC85 model equation (\ref{winddensity}), assuming the mass fraction of hydrogen for solar abundances $X_{\rm H}\approx0.71$. The ionization fraction $n_{e}/n_{\rm H}$ is also calculated by SPEX (\citealt{Schure09}). Thus, the total X-ray luminosity between two frequencies $\nu_{1}$ and $\nu_{2}$ is written as
\begin{eqnarray}
L_{X,\rm hot}^{[\nu_{1},\nu_{2}]}&\approx&1.5\times10^{8}L_{\odot}\;X_{\rm H}^{2}\left(\frac{\beta^{3}}{\alpha}\frac{\textrm{SFR}_{0}^{2}}{R_{200\rm pc}}\right)\nonumber\\
&&\times\int_{0}^{\infty}dr_{*}\;r_{*}^{2}\rho_{*}^{2}\Lambda_{\rm N,-22}^{[\nu_{1},\nu_{2}]}\left(\frac{n_{e}}{n}\right),\label{Lxhot}
\end{eqnarray}
where $\Lambda_{N,-22}^{[\nu_{1},\nu_{2}]}=\Lambda_{\rm N}^{[\nu_{1},\nu_{2}]}/10^{-22}$ erg s$^{-1}$ cm$^{3}$, other variables are denoted in Section \ref{sectCC85model}. Note the scaling of $L_{X,\rm hot}$ with SFR, and that equation (\ref{winddensity}) gives $n\propto$ SFR. Thus equation (\ref{Lxhot}) shows that X-ray emission from the hot wind fluid scales as
\begin{equation}
L_{X,\rm hot}\propto n^{2}\Lambda_{\rm N}\propto \textrm{SFR}^{2},
\end{equation}
significantly steeper than the observed linear relation $L_{X}\propto$ SFR for star-forming galaxies.

\section{The Observed $L_{X}-$SFR Relation Constrains Hot Winds}\label{sect_LxSFR}

In this paper we focus on the diffuse X-ray emission from the hot wind fluid.  Because the fraction of the observed hard X-ray emission that is actually due to diffuse gas rather than other sources is uncertain, we adopt the following relation between the diffuse emission at $0.5-8$ keV and SFR based on the discussion in Section \ref{Intro}:
\begin{equation}
L_{X,\rm diffuse\;(0.5-8\,\rm keV)}=4.0\times10^{39}\;f_{d}\;{\rm erg\;s}^{-1}\textrm{SFR}/(M_{\odot}\;{\rm yr}^{-1}),\label{diffuse}
\end{equation}
where $f_{d}\leq1$ is the fraction of the diffuse emission in X-rays due to the hot wind fluid, and $f_{d}=1$ is the observed mean relation between total X-ray emission and SFR. In general, we expect $f_{d}\sim0.1$, as seen in M82 (\citealt{SH09}), but higher and lower values are considered throughout this paper. In Section \ref{discussion} we discuss constraints on $f_{d}$ based on some well-studied starbursts. Theoretical constraints on the hot wind fluid from the observed diffuse X-ray emission can then be obtained by combining equations (\ref{Lxhot}) and (\ref{diffuse}) such that
\begin{equation}
L_{X,\rm hot}=L_{X,\rm diffuse},\label{constraint}
\end{equation}
where the hard X-ray emission is from 0.5 keV to 8 keV (i.e., $\nu_{1}=0.5$ keV and $\nu_{2}=8$ keV in equation [\ref{Lxhot}] and Fig. \ref{fig_Cooling}).

%------------------------------------------------------------------------

\begin{figure}[t]
\centerline{\includegraphics[width=10cm]{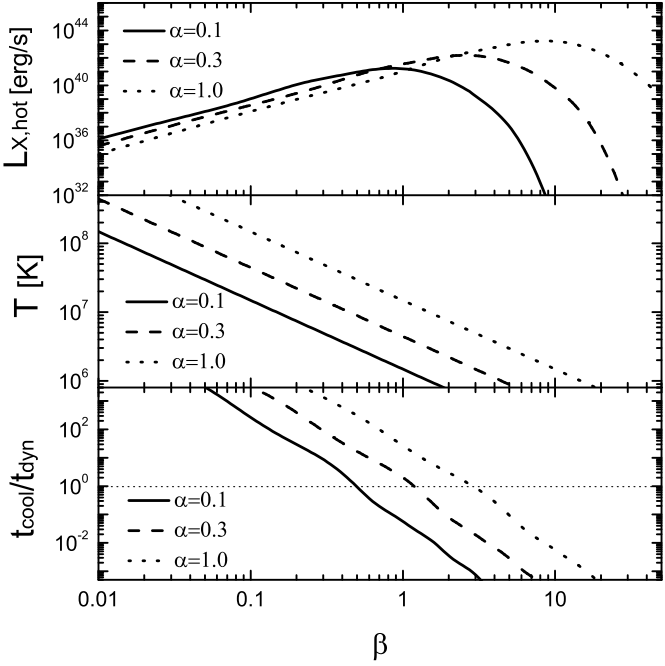}}
\caption{Total hot X-ray luminosity $L_{X,\rm hot}$ between 0.5-8 keV (upper), galactic center temperature $T(r=0)$ (middle) and the ratio of the cooling and dynamic timescale $t_{\rm cool}/t_{\rm dyn}(r=R)$ (lower) as a function of $\beta$ for fixed $\alpha$. The wind launching radius $R=200$ pc and SFR $=10\,M_{\odot}$ yr$^{-1}$.}\label{fig_Lxhot}
\end{figure}
%------------------------------------------------------------------------%

%------------------------------------------------------------------------%
\begin{figure*}[t]
\centerline{\includegraphics[width=18cm]{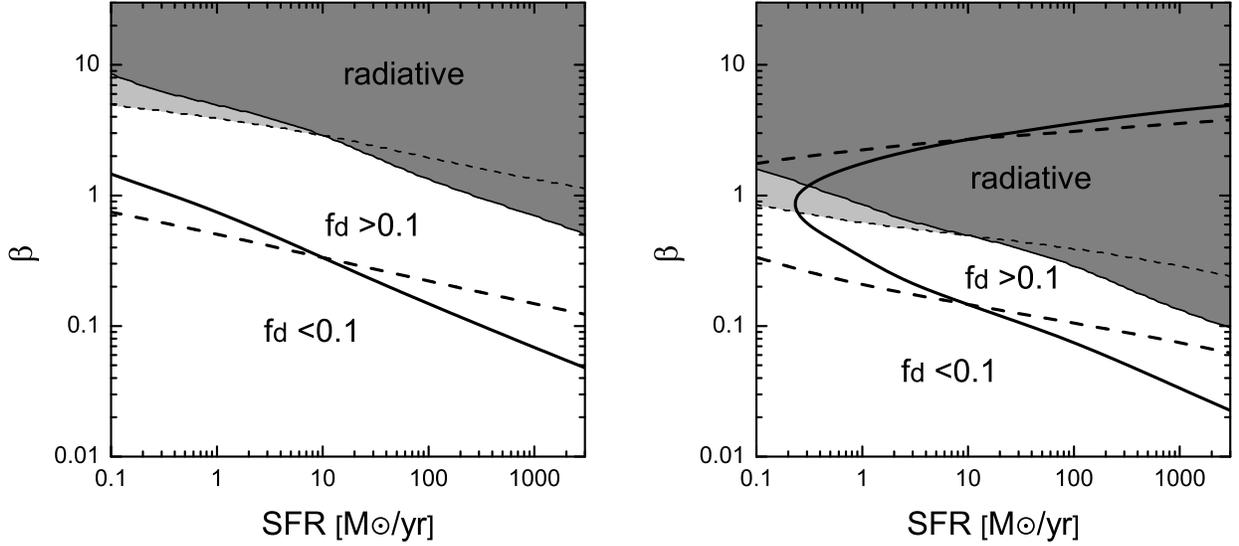}}
\caption{Solutions of $\beta$ as a function of SFR for fixed $\alpha=1$ (left panel) and $\alpha=0.1$ (right panel), with $f_{d}=0.1$, and the wind launching radius R=200 pc (solid lines), $R=200$ (SFR/SFR$_{\rm n})^{1/2}$ pc (dashed lines) with the normalization SFR$_{\rm n}$=10 $M_{\odot}$ yr$^{-1}$ (dotted lines). The gray regions give where the flow is radiative $t_{\rm dyn}\geq t_{\rm cool}$ at $r=R$.}\label{fig_SFR1}
\end{figure*}
%------------------------------------------------------------------------%

%------------------------------------------------------------------------%
\begin{figure*}[t]
\centerline{\includegraphics[width=18cm]{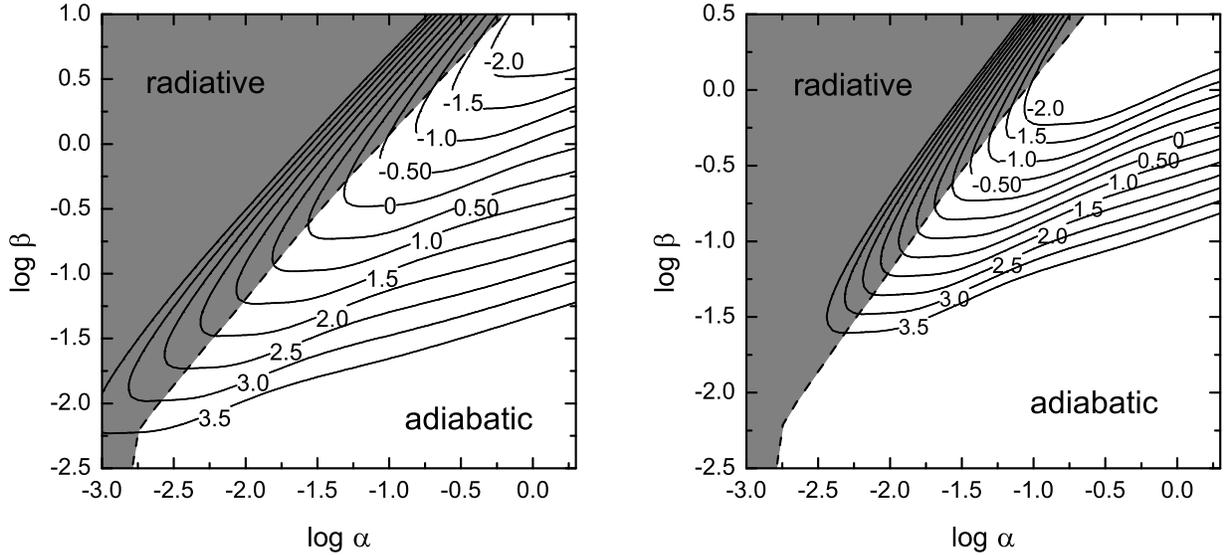}}
\caption{Contours of solutions of SFR $\log_{10}$(SFR$/M_{\odot}$ yr$^{-1}$) as a function of $(\alpha,\beta)$ by equation (\ref{constraint}), with $f_{d}=0.1$ and $R=200$ pc (left) and $R=200$ (SFR/SFR$_{\rm n})^{1/2}$ with SFR$_{\rm n}$=10 $M_{\odot}$ yr$^{-1}$ (right). The dashed line is the critical line $t_{\rm dyn}=t_{\rm cool}$ at $r=R$ based on equation (\ref{betaconstraint2}).}\label{fig_SFR2}
\end{figure*}
%------------------------------------------------------------------------%

%------------------------------------------------------------------------%
%\begin{figure*}[p]
\begin{figure*}[t]
\centerline{\includegraphics[width=18cm]{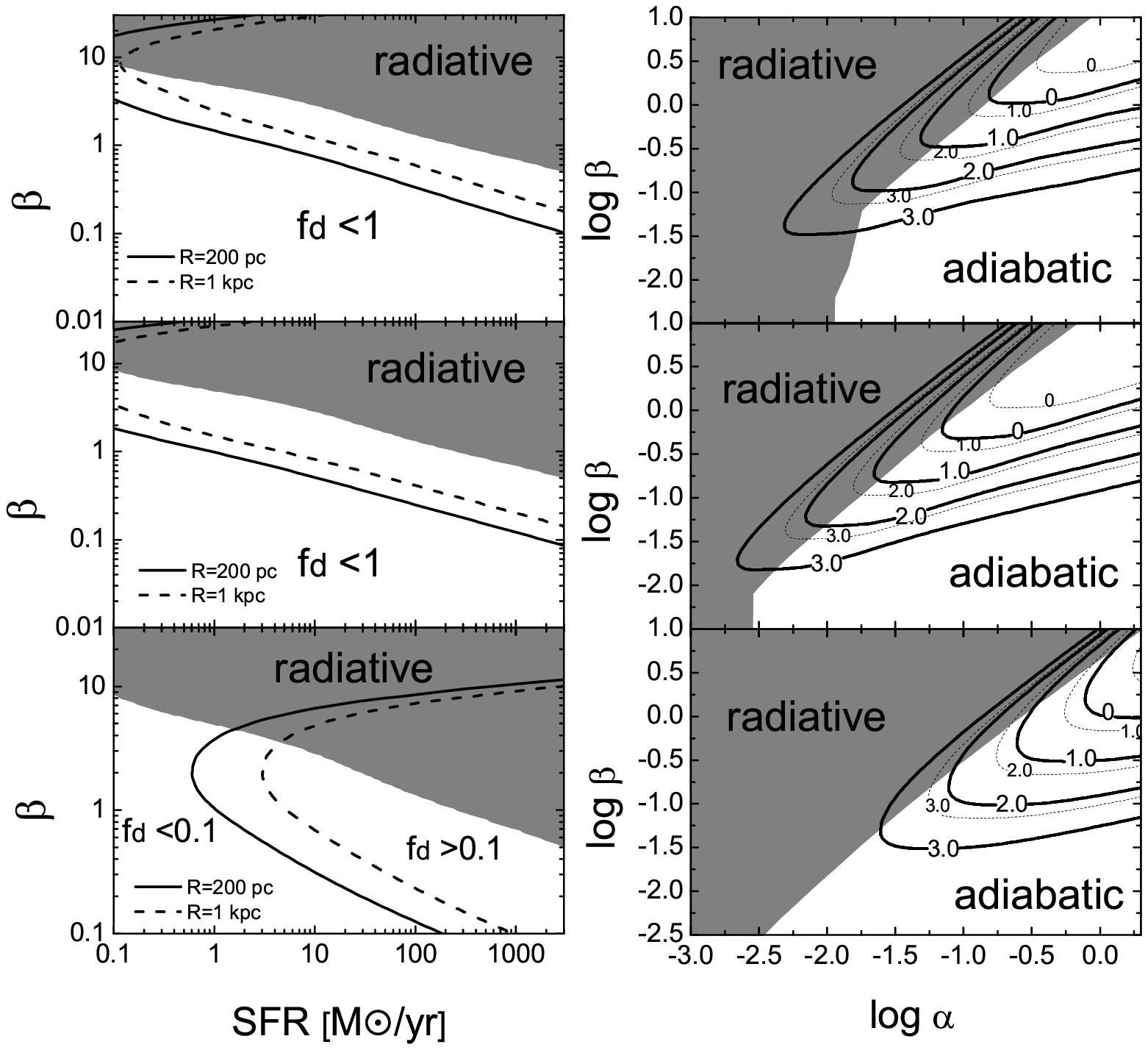}}
\caption{Solutions for $\beta$ as a function of SFR for fixed $\alpha=1$ (left panels) and contours of SFR $\log_{10}$(SFR$/M_{\odot}$ yr$^{-1}$) as a function of $\alpha$ and $\beta$ (right panels) for different $R$: $R=200$ pc (solid lines) and 1 kpc (dashed lines). The gray radiative regions in left panels are excluded by equation (\ref{betaconstraint}) with $R=200$ pc, while in right panels the gray radiative regions are excluded by equation (\ref{betaconstraint2}) with the corresponding $f_{d}$ mentioned below. The upper panels show the results based on the soft X-ray $L_{X}-$SFR relation using equation (\ref{diffuse}) with $f_{d}=1$, the middle panels are based on $L_{X}-$SFR relation in \cite{Mineo12b} with $f_{d}=1$, and the lower panels are based on the hard X-ray $L_{X}-$SFR relation in \cite{Lehmer10} with $f_{d}=0.1$.}\label{fig_discuss}
\end{figure*}
%------------------------------------------------------------------------%

There are five parameters in equation (\ref{constraint}): $\alpha$, $\beta$, SFR, $R$, and $f_{d}$. Using equation (\ref{constraint}), the relation between any two of the five parameters can be constrained by the other three parameters. However, the constraints on the CC85 model, in particular, the parameter set of $(\alpha,\beta)$ are of particular interest, since $\alpha$ and $\beta$ determine the importance of the hot gas for driving matter out of the galaxy directly, and for driving swept up cold gas clouds out of the galaxy via ram pressure. If we fix $f_{d}$ in equation (\ref{constraint}), with a fixed $R$ (or apply an independent model of $R({\rm SFR})$), we can solve for the relation between $\alpha$, $\beta$, and SFR.

Figure \ref{fig_Lx} shows X-ray luminosities $L_{X,\rm hot}$ as a function of SFR with some given values of $(\alpha,\beta)$ (black lines), compared with the diffuse luminosity $L_{X,\rm diffuse}$ with different $f_{d}$ (blue lines). The radius $R$ is fixed at $R=200$ pc for illustrative purposes. Other values of $R$ give similar results. The slope of $L_{X,\rm hot}$ is steeper than $L_{X,\rm diffuse}$, due to the different scaling between bremsstrahlung emission from the hot wind $L_{X,\rm hot}\propto$ SFR$^{2}$, and the observed X-ray luminosity $L_{X}\propto$ SFR (equations [\ref{Lxhot}] and [\ref{diffuse}]). Each point of intersection between two lines $L_{X,\rm hot}$ and $L_{X,\rm diffuse}$ for any value of $f_{d}$ gives a certain parameter set of $\alpha$, $\beta$ and SFR. For example, for $f_{d}=0.1$ and $(\alpha,\beta)=(1,1)$, Figure \ref{fig_Lx} shows that $L_{X,\rm hot}$ intersects $L_{X,\rm diffuse}$ at SFR $\sim0.4$\,M$_{\odot}$ yr$^{-1}$, implying that for $f_{d}=0.1$, $(\alpha,\beta)=(1,1)$ is only allowed for galaxies with SFR $\lesssim0.4$\,M$_{\odot}$ yr$^{-1}$; for SFR $\gtrsim0.4$\,M$_{\odot}$ yr$^{-1}$, $(\alpha,\beta)=(1,1)$ produces too much diffuse X-ray emission. Similarly, note that $(\alpha,\beta)=(1,10)$ is completely ruled out because it always produces too much diffuse X-ray emission. One might expect that in general higher values of $\beta$ (i.e., larger mass loading) would always lead to higher diffuse X-ray luminosity and tighter constraints on the allowed range of SFR, but the $L_{X, \rm hot}$ lines for $\alpha=0.1$ (thin lines) show that this is not the case. In fact, $L_{X,\rm diffuse}$ for $(\alpha,\beta)=(0.1, 5)$ falls below other lines with lower $\beta=0.1$ and 1, this is because the temperature of the flow for $(\alpha,\beta)=(0.1,5)$ is so cool that only a small amount of the emission from the wind fluid is in the X-ray band. In fact, multiple solutions for $\beta$ exist at fixed $\alpha$ and SFR. For example, for $f_{d}=0.1$ and $\alpha=0.1$, both $\beta\simeq0.1$, and $\beta\sim 1$ are valid solutions in the CC85 model for ${\rm SFR}=10$\,M$_{\odot}$ yr$^{-1}$.

Figure \ref{fig_Lxhot} makes the solutions for $\beta$ more explicit. For fixed $\alpha$, $L_{X,\rm hot}$ in equation (\ref{Lxhot}) has a maximum as a function of $\beta$. The top panel of Figure \ref{fig_Lxhot} shows that $L_{X,\rm hot}$ is thereby peaked as a function of $\beta$. For a given $L_{X,\rm hot}$ below the peak, there are two solutions for $\beta$, one low, and another high. As shown in the middle panel of Figure \ref{fig_Lxhot}, the low value of $\beta$ corresponds to high $T$, and high $\beta$ to low $T$. However, not all solutions of $\beta$ are physically realizable in the CC85 model, because of radiative cooling. The CC85 model assumes that the flow is adiabatic for $r>R$ and fluid cooling is not important throughout the wind profile. If the cooling timescale in a mass loaded wind $t_{\rm cool}$ becomes smaller than the local dynamical timescale $t_{\rm dyn}\sim r/V_{\rm hot}$, then this assumption is invalidated and the CC85 model must break down. The wind dynamical timescale is
\begin{equation}
t_{\rm dyn}\sim2.8\times10^{5}\,\textrm{yr}\;\alpha^{-1/2}\beta^{1/2}u_{*}^{-1}R_{200pc}\left(\frac{r}{R}\right).
\end{equation}
The total energy is $\varepsilon_{\rm heat}=\rho\left(\frac{1}{2}V_{\rm hot}^{2}+\frac{c_{s}^{2}}{\gamma-1}\right)=\rho_{*}\dot{E}^{1/2}\dot{M}^{1/2}/R^{2}$, thus the cooling timescale is estimated by $t_{\rm cool}\sim \varepsilon_{\rm heat}/(n_{e}n_{\rm H}\Lambda_{\rm N})$. Note that the ratio of $t_{\rm cool}/t_{\rm dyn}$ of the hot wind has its lowest value at $r=R$; thus we can focus on $t_{\rm cool}/t_{\rm dyn}$ at $r=R$ as the strongest timescale constraint. If we take bremsstrahlung emission from the pure hydrogen gas as a lower limit for the cooling rate $\Lambda_{\rm N}$, we obtain an analytic upper bound on $\beta$:
\begin{equation}
\beta\leq6.6\,\alpha^{3/5}R_{200pc}^{2/5}\left(\frac{10\,M_{\odot}\,\textrm{yr}^{-1}}{\rm SFR}\right)^{2/5}\left(\frac{\Lambda_{\rm brems}^{\rm H}}{\Lambda_{\rm N}}\right)^{2/5}.\label{betaconstraint}
\end{equation}
The lower panel of Figure \ref{fig_Lxhot} shows the numerical results for $t_{\rm cool}/t_{\rm dyn}$ at $r=R$, which decreases strongly with increasing $\beta$. Since $t_{\rm cool}<t_{\rm dyn}$ at high $\beta$, these high-$\beta$ solutions are not physical in CC85 adiabatic wind model. 

Figure \ref{fig_SFR1} shows both the solution for $\beta$ from equation (13) as a function of SFR, and constraints from radiative cooling. The thick lines in each panel are solutions for fixed $f_{d}=0.1$ for two different models of $R$: $R=200$ pc and $R\propto$ SFR$^{1/2}$ (\citealt{Heckman00}; see also \citealt{Lehnert96}, \citealt{Meurer97}, \citealt{Martin05}), and fixed $\alpha=1$ (left panel) and 0.1 (right panel). The thin solid and dashed lines show the critical condition $t_{\rm cool}=t_{\rm dyn}$ at $r=R$ from equation (\ref{betaconstraint}), where we have used the total emissivity $\Lambda_{N}$ from Figure \ref{fig_Cooling}. Above the line $t_{\rm cool}<t_{\rm dyn}$, the wind is radiative outside the host galaxy, and the CC85 model breaks down. For $\alpha=0.1$, there are two solutions for $\beta$, but the high-$\beta$ solutions are excluded by the radiative cooling constraint (equation [\ref{betaconstraint}]). Below the solution curves $f_{d}<0.1$, and above these curves $f_{d}>0.1$. Importantly, one finds that compact rapidly star-forming galaxies should in general have relatively low $\beta$. Taking $\alpha=1$ and $R=200$ pc, we see that for ${\rm SFR}\gtrsim10$\,M$_\odot$ yr$^{-1}$ that $\beta\lesssim0.3$. For lower $\alpha=0.1$, the limit on $\beta$ decreases to $\beta\lesssim0.15$. Thus, it does not appear possible for the CC85 alone to account for $\beta\sim1-10$, as seems to be required in galaxy formation models \citep{Springel03,Oppenheimer_Dave06,Oppenheimer_Dave08,Finlator_Dave08,Bower12,Puchwein13}. For example, \cite{Puchwein13} (see also \citealt{Bower12}) shows that $\beta\sim1-10$ is required to match the observed galaxy stellar mass function. Note that for lower assumed values of $\alpha$ or $f_d$ all of the solution curves move down in $\beta$, making the limits on $\beta$ even stronger.

Constraints on $\beta$ for a broader range of $\alpha$ are showed in the $\alpha-\beta$ plane in Figure \ref{fig_SFR2}, which shows contours of solutions of $\log_{10}($SFR$/M_{\odot}$ yr$^{-1})$ by equation (\ref{constraint}), with fixed $f_{d}=0.1$, and $R=200$ pc and $R\propto$ SFR$^{1/2}$. For example, taking SFR$=10\,M_{\odot}$ yr$^{-1}$, we see that $\beta\lesssim0.3$ for $\alpha=1$ and $\beta\lesssim0.06$ for $\alpha=0.01$. Also, we find that larger $R$ yields a smaller allowed $(\alpha,\beta)$ space for the same SFRs. Note that the cooling constraint equation (\ref{betaconstraint}) is independent of equation (\ref{constraint}). Combining equations (\ref{constraint}) and (\ref{betaconstraint}) by solving SFR by equation (\ref{constraint}) as a function of $\alpha$, $\beta$, $R$ and $f_{d}$, the cooling constraint becomes a function of $f_{d}$. Thus $t_{\rm cool}/t_{\rm dyn}(r=R)\geq1$ takes the form
\begin{eqnarray}
&&6.3\times10^{3}\alpha\Lambda_{\rm N,-22}^{-1}f_{d}^{-1}\left(\frac{n_{e}}{n_{\rm H}}\right)^{-1}\nonumber\\
&&\times\int_{0}^{\infty}dr_{*}\;r_{*}^{2}\rho_{*}^{2}\Lambda_{\rm N,-22}^{[\nu_{1},\nu_{2}]}\left(\frac{n_{e}}{n_{\rm H}}\right)\geq1,\label{betaconstraint2}
\end{eqnarray}
which is independent of $R$. The grey region above the thick solid diagonal line ($t_{\rm cool}=t_{\rm dyn}$ at $r=R$) in Figure \ref{fig_SFR2} denotes radiative winds excluded by equation (\ref{betaconstraint2}). The allowed values of $\beta$ are seen to decrease rapidly for lower $\alpha$ and SFR at fixed $f_{d}$.

Finally, we note that \cite{CC85} ignored the effect of gravity, because they focused on solutions for M82 with $\alpha\approx\beta$, and thus they derived a terminal velocity of the M82 outflow of $V_{\rm term}\simeq1000$ km s$^{-1}$, much higher than the escape velocity from the galaxy $V_{\rm esc}$ of a few hundred km s$^{-1}$ \citep{greco}. If we consider the general constraint that requires $V_{\rm term}>V_{\rm esc}$ for an unbound outflow we find that
\begin{equation}
\beta<\alpha\left(\frac{V_{\rm esc}}{1000\;\textrm{km}\;\textrm{s}^{-1}}\right)^{-2}.
\end{equation}
For $V_{\rm esc}=200$ km s$^{-1}$, this criterion is not restrictive, but for deep gravitational potentials ($V_{\rm esc}\gtrsim400$ km s$^{-1}$), this limit on $\beta$ becomes close to the constraints produced by the critical lines of $t_{\rm cool}=t_{\rm dyn}$ in Figures \ref{fig_SFR1} and \ref{fig_SFR2}.

\section{Discussion}\label{discussion}

There are several parameters in our model for the difuse X-ray emission associated with galactic winds, including the diffuse fraction $f_{d}$, the normalization factor for the observed $L_{X}-$SFR correlation, and the selected X-ray energy range. It is important to investigate whether our constraints on $\alpha$ and $\beta$ change with changing parameters in the model. 

Constraints on $f_{d}$ can also be given by some well-studied starbursts with diffuse X-ray data, such as M82, NGC 253, NGC 6240 and Arp 220. We use equation (\ref{diffuse}) to calculate $f_{d}$. Figure \ref{fig_Lx} shows the diffuse X-ray emission from these galaxies in blue diamonds. The observed extended X-ray luminosity in M82 from 0.1 to 2.4 keV band is $L_{X}^{0.1-2.4 \rm keV}\sim1.9\times10^{40}$ erg s$^{-1}$ (\citealt{Strickland97}), and in the 2-8 keV band is $L_{X}^{2-8 \rm keV}\leq 4.4\times10^{39}$ erg s$^{-1}$ (\citealt{SH07}). The total 8-1000 $\mu$m infrared luminosity of M82 $L_{\rm IR}\simeq5.6\times10^{10}L_{\odot}$ (\citealt{Sanders03}) corresponds to a SFR of $\sim5-10\,M_{\odot}$ yr$^{-1}$ (\citealt{OConnell78}; \citealt{Kennicutt98}; \citealt{FS03}; \citealt{Strickland04}; \citealt{Elbaz07}; \citealt{SH09}; \citealt{Panuzzo10}), depending on the assumed IMF. If we convert the observed diffuse X-ray from $2-8$ keV to $0.5-8$ keV with a conversion factor of 1.5, as assumed in \cite{Mineo12a}, we get the diffuse fraction $f_{d}\lesssim0.16$ for the upper bound of SFR as 10 $M_{\odot}$ yr$^{-1}$, and $f_{d}\lesssim0.32$ for a lower estimated SFR $=5\,M_{\odot}$ yr$^{-1}$.  Another starburst with diffuse X-ray data is NGC 253, which has an unabsorbed $2-10$ keV luminosity of $2\times10^{39}$ erg s$^{-1}$ (\citealt{Weaver02}). Taking the total SFR of NGC 253 as 5 $M_{\odot}$ yr$^{-1}$ (\citealt{Melo02}), we get $f_{d}\sim0.13$. NGC 6240, however, has been recently observed with a high diffuse luminosity $5.3\times10^{41}$ erg s$^{-1}$ at $0.5-8$ keV (\citealt{Wang13}), which indicates a high value of $f_{d}\simeq1.3$ for a total SFR of $\sim100$ $M_{\odot}$ yr$^{-1}$ (\citealt{Heckman00}). A higher estimated SFR of $\sim200$ $M_{\odot}$ yr$^{-1}$ gives $f_{d}\sim0.7$. \cite{McDowell03} observed individual sources of extended X-ray emission in Arp 220. Following their suggestion that the inner two plumes of hot gas with a luminosity of $\sim3\times10^{40}$ erg s$^{-1}$ are due to a galactic wind, we get $f_{d}\sim0.03$ for Arp 220. In short, the nearby well-studied starbursts show that in general it is reasonable to take $f_{d}\lesssim0.1$ as the typical value for the diffuse X-ray emission. However, for some galaxies, such as NGC 6240, $f_{d}\sim1$ is also possible. The upper panels in Figure \ref{fig_discuss} shows solutions for $\beta$ using equations (\ref{constraint}) and the $L_{X}-$SFR relation (eq. [\ref{diffuse}]), but for $f_{d}=1$, $R=200$ pc and 1 kpc. Compared with Figures \ref{fig_SFR1} and \ref{fig_SFR2} (left panels), higher $f_{d}$ or larger $R$ do not change our results qualitatively. For example, for $\alpha=1$ we still find $\beta\lesssim1$ for ${\rm SFR}\gtrsim10$\,M$_\odot$ yr$^{-1}$.

We also check whether different normalizations of $L_{X}-$SFR based on a different SFR estimate or energy range change our results. For example, \cite{Mineo12b} studied the soft X-ray luminosity of the diffuse ISM and obtained the linear relation
\begin{equation}
L_{X\,(0.5-2)\,\rm{keV}}^{\rm tot}/\textrm{SFR}\simeq(8.3\pm0.1)\times10^{38}\;\textrm{erg}\,\textrm{s}^{-1}/(M_{\odot}\,\textrm{yr}^{-1}),
\label{Lxsoft}
\end{equation}
over the range of SFRs from $\sim0.1$ to $\sim17$ $M_{\odot}$ yr$^{-1}$. Assuming equation (\ref{Lxsoft}) holds for all SFRs, the middle panels of Figure \ref{fig_discuss} give constraints on $\beta$ with a maximum $f_{d}=1$, which assumes all soft X-rays are from the diffuse component. As another example, we adopt the hard X-ray $L_{X}-$SFR scaling by \cite{Lehmer10} (their Table 4, Model 1)
\begin{equation}
L_{X\,(2-10)\,\rm{keV}}^{\rm tot}/\textrm{SFR}\simeq10^{39.24\pm0.06}\;\textrm{erg}\,\textrm{s}^{-1}/(M_{\odot}\,\textrm{yr}^{-1}).
\label{Lxhard2}
\end{equation}
Taking the typical value $f_{d}=0.1$, the lower panels in Figure \ref{fig_discuss} show the results using Equation (\ref{Lxhard2}) with $R=200$ pc and 1 kpc. In general, for $\alpha=1$, we always find that $\beta\lesssim1$ for SFR$>10\,M_{\odot}$ yr$^{-1}$, with an uncertainty of a factor of a few for different $L_{X}-$SFR normalizations, and adopted values of $f_{d}$ and $R$. The middle and lower contour panels in Figure \ref{fig_discuss} show that the hard X-ray constraint is weaker than the soft X-ray constraints from equation (\ref{Lxsoft}), but our basic conclusion that large $\beta$ is excluded in the CC85 model is not changed qualitatively. 

It is worthwhile to highlight the limitations of the CC85 model. Note that the breakdown of the adiabatic CC85 model at large $\beta$ as  a result of radiative cooling highlighted in Section \ref{sect_LxSFR} is consistent with \cite{Silich03,Silich04}, who showed that an adiabatic stationary solution of the wind does not exist with $\dot{E}_{\rm hot}$ or $\dot{M}_{\rm hot}$ larger than a critical value. Too large $\dot{M}_{\rm hot}$ produces a so-called bimodal solution of the flow (e.g., \citealt{Tenorio-Tagle07}; \citealt{Wunsch08}; \citealt{Wunsch11}), in which the densest inner regions immediately radiate away the deposited energy while the outer zones develop a strongly radiative wind. We compute the ratio of $n_{e}n_{\rm H}\Lambda_{\rm N}/q_{\rm hot}$, where $q_{\rm hot}=3\dot{E}_{\rm hot}/(4\pi R^{3})$ is the volumetric heating rate inside the galaxy. Since $n_{e}n_{\rm H}\Lambda_{\rm N}/q_{\rm hot}$ decreases as a function of radius inside the galaxy, we focus on the ratio $n_{e}n_{\rm H}\Lambda_{\rm N}/q_{\rm hot}$ at the center of the host galaxy ($r=0$). If $n_{e}n_{\rm H}\Lambda_{\rm N}/q_{\rm hot}>1$ at the center, the hot fluid becomes radiative and unstable as discussed by \cite{Silich04} and their series of papers, and again the assumptions of the CC85 model break down. Although the thick solid line in Figure \ref{fig_SFR2} corresponds to $t_{\rm cool}=t_{\rm dyn}$, we find that it gives a similar bound to $n_{e}n_{\rm H}\Lambda_{\rm N}=q_{e}$ at $r=0$, so that for $\beta$ above this line we expect an unstable and radiative hot fluid inside the galaxy.

Moreover, the CC85 model also has some other simplifications. In particular, it assumes spherical symmetry and radius-independent energy thermalization and mass-loading efficiency densities (i.e., constant $q$ and $Q$ in the CC85 model, see Appendix \ref{CC85appendix}). The issue of spherical symmetry has been addressed by \cite{SH09}, who find that it is possible to use the spherical analytical CC85 model to predict the wind fluid properties and X-ray emission from a disklike starburst for M82-like systems. Similarly, one might imagine that a more realistic volumetric energy thermalization and mass-loading efficiency densities that are functions of radius would change the properties of the wind solutions and call our results into question. For simplicity if we assume a radius-dependent energy and mass injection density $Q\propto r^{-\xi}$ and $q\propto r^{-\xi}$ in the CC85 model, the solutions of the hot wind ($u_{*}$, $\rho_{*}$ and $P_{*}$) change inside the galaxy $r<R$ (see equations [\ref{powerlaw1}], [\ref{powerlaw2}] and [\ref{dimenless24}] in Appendix \ref{CC85appendix} as the solutions). For example, the Mach number of the wind becomes $M\approx \sqrt{(\xi-1)/[\gamma(3-\xi)]}$ at $r\ll R$ for $1<\xi<3$, which is different from the case of $\xi<1$ that $M\approx 0$ at $r\ll R$. For the purpose of this paper we focus on integrated constraints $\alpha$ and $\beta$ given by equations (\ref{parameter2}) and (\ref{parameter3}). For example, we find that for $\alpha=1$, the constraint on $\beta$ for SFR $\gtrsim1\;M_{\odot}$ yr$^{-1}$ changes by less than a factor of 50\% as a function of $\xi$ in the range of $0\leq \xi <3$. For a wide range of $\alpha\lesssim2$, the basic conclusion in this paper that $\beta\lesssim1$ for SFR$\gtrsim10\,M_{\odot}$ yr$^{-1}$ in the CC85 model still holds. Of course real galaxies are clumpy, with a complex ISM. It is worthwhile to study the effects of multiphase and clumpy nature of wind on the dynamics and X-ray emission of the wind. Some hydrodynamical simulations with more realistic and complex wind structure (e.g., \citealt{Cooper08,Cooper09}; \citealt{Fujita09}, \citealt{hopkins_wind}) showed comparisons between X-ray emission in the model and observations, as well as estimate on mass-loading efficiency. However, these works are mostly applied for M82-like and dwarf starbursts. More detailed X-ray comparisions between models and observations should be done for starburst galaxies across a wide range of SFRs in the future.

\section{Conclusions}\label{conclusions}

In this paper we constrain the properties of any potential hot galactic wind from galaxies varying from dwarf starbursts to ULIRGs using the adiabatic hot wind model of \cite{CC85}. Numerical simulations have shown that this model provides a good description of the hot wind fluid (e.g., \citealt{SH09}). We use the observed total X-ray luminosities of galaxies to constrain the efficiency  $\alpha$ (i.e., energy input rate; eq.~\ref{parameter2}) and the mass-loading efficiency $\beta$ (eq.~\ref{parameter3}) of the hot wind.  We first showed that the diffuse hard X-ray luminosity from any putative hot outflow should scale as $L_{X,\rm hot}\propto$ SFR$^{2}$ for fixed $\alpha$ and $\beta$ (eq.~\ref{Lxhot}).  However, observations show a linear relation between the total galactic X-ray luminosity and SFR (Fig.~\ref{fig_Lx}). If we attribute a fraction ($f_d<1$) of this total emission  to the hot wind fluid we derive constraints on $\alpha$ and $\beta$ as a function of SFR and the wind launching radius $R$ (Figs.~\ref{fig_SFR1} and \ref{fig_SFR2}).  We showed that for fixed $f_d$ there exist multiple solutions for $\beta$ (one low and one high) as a function of $\alpha$ and SFR (Fig.~\ref{fig_Lxhot}).  We highlight the importance of radiative cooling for the heavily mass-loaded high-$\beta$ solutions, which invalidates the adiabatic CC85 model (eq. \ref{betaconstraint}). The breakdown of the CC85 model at large $\beta$ is consistent with \cite{Silich03, Silich04}, who showed that the adiabatic stationary solution of the wind does not exist with $\dot{E}_{\rm hot}$ or $\dot{M}_{\rm hot}$ larger than a critical value. As a result, we showed that only moderate mass-loading is allowed in the CC85 model. For example, as Figure \ref{fig_SFR2} gives, for reasonable values of the thermalization efficiency $\alpha\lesssim1$, and for ${\rm SFR}\gtrsim10$\,M$_\odot$ yr$^{-1}$ we find that $\beta\lesssim1$; higher values of $\beta$ would require that significantly more of the total X-ray luminosity of star-forming galaxies be attributable to the hot flow, in conflict with observations that find that the X-ray luminosities of galaxies are dominated by compact objects. These conclusion do not change for reasonable variations about our fiducial model (Fig. \ref{fig_discuss}). 

Our result that $\beta\lesssim1$ for SFR$\gtrsim10\,M_{\odot}$ yr$^{-1}$ shows that it does not appear possible for an adiabatic CC85 model alone to account for $\beta\sim1-10$ as seems to be required by integrated constraints on the efficiency of stellar feedback in galaxies \citep{Springel03,Oppenheimer_Dave06,Oppenheimer_Dave08,Finlator_Dave08,Bower12,Puchwein13}. For example, \cite{Puchwein13} (see also \citealt{Bower12}) show that $\beta\sim1-10$ is required to match the observed galaxy stellar mass function. However, it remains possible that the swept up cold gas carries most of the mass (e.g., \citealt{Cooper08, Cooper09, Fujita09}), or that other wind driving mechanisms such as momentum deposition by supernovae and radiation pressure on dust (\citealt{MQT05}) or cosmic rays (\citealt{Everett08, Socrates08}) dominate wind driving.

Finally, there are several additional studies that can further constrain and explain hot winds. First, since the CC85 model assumes a radius-independent energy input and mass-loading efficiency densities in the wind, it is worthwhile to investigate the effects of multiphase and clumpy nature of winds on their dynamics and X-ray emission across a wide range of galactic SFRs. We encourage more detailed multi-dimensional simulations like those of \cite{Cooper08,Cooper09}, \cite{Fujita09} and \cite{hopkins_wind}, but for massive starbursts like Arp220 and other ULIRGs to assess their integrated X-ray emission for comparison with observations. Second, the inclusion of a model of cold cloud acceleration and destruction in the CC85 model may further constrain the hot wind model. Note that a traditional way to explain the cold gas in galactic winds is that the cold gas clouds enter the SN-heated hot wind flow at relatively low velocity and are accelerated by the ram pressure of the hot wind. In a future work, we will combine a model of cold cloud acceleration and destruction with the CC85 model to constrain energy thermalization and hot-wind mass-loading efficiency in individual systems that have measured cold cloud velocities, incorporating the constraints on the hot wind determined in this paper. Lastly, it is clear that radiative cooling is important for the dynamics of outflows across a wide range of parameter space of thermalization input and high mass-loading efficiency. Formulating radiative solutions for high$-\beta$ winds and comparing with observations is an important direction for future work.

\section{Acknowledgements}
We thank the anonymous referee for comments that have allowed us to improve our paper. D.Z. and T.A.T. thank Smita Mathur and David Weinberg for helpful discussions on the cooling functions of X-ray emission and wind mechanisms. T.A.T. also thanks Tim Heckman for a number of stimulating discussions. This work is supported in part by NASA grant \# NNX10AD01G. E.Q. is supported in part by NASA ATP Grant 12-ATP12-0183, a Simons Investigator award from the Simons Foundation, the David and Lucile Packard Foundation, and the Thomas Alison Schneider Chair in Physics. N.M. is supported in part by NSERC of Canada and by the Canada Research Chair program.

\appendix
\section{CC85 model}\label{CC85appendix}

CC85 present a spherical thermal winds where gravitational forces can be ignored. Assuming inside the radius of the starburst region $R$ hte total mass and energy input are $\dot{M}_{\rm hot}$ and $\dot{E}_{\rm hot}$ respectively, with an averaged injected efficiencies per unit volume in the galaxy being $q=\dot{M}_{\rm hot}/V$, $Q=\dot{E}_{\rm hot}/V$ and $V=4\pi R^{3}/3$. The one-dimensional hydrodynamic equations for the hod fluid are
\begin{eqnarray}
\frac{1}{r^{2}}\frac{d}{dr}\left(\rho u r^{2}\right)=q\\
\rho u \frac{du}{dr}=-\frac{dP}{dr}-qu\\
\frac{1}{r^{2}}\frac{d}{dr}\left[\rho u r^{2}\left(\frac{1}{2}u^{2}+\frac{\gamma}{\gamma-1}\frac{P}{\rho}\right)\right]=Q.
\end{eqnarray}
The solutions for the Mach number $M$ of the hot wind fluid are given in CC85 by (see also \citealt{Canto00})
\begin{equation}
\left[\frac{(\gamma-1)+2/M^{2}}{\gamma+1}\right]^{(1+\gamma)/[2(1+5\gamma)]}
\left[\frac{3\gamma+1/M^{2}}{1+3\gamma}\right]^{-(3\gamma+1)/(5\gamma+1)}=\frac{r}{R},
\end{equation}
for $r<R$ and
\begin{equation}
M^{2/(\gamma-1)}\left(\frac{\gamma-1+2/M^{2}}{1+\gamma}\right)^{(\gamma+1)/[2(\gamma-1)]}=\left(\frac{r}{R}\right)^{2}.
\end{equation}
for $r>R$. We take the dimensionless variables as
\begin{eqnarray}
&&P=P_{*}\dot{M}^{1/2}\dot{E}^{1/2}R^{-2}\\\label{AppenP}
&&\rho=\rho_{*}\dot{M}^{3/2}\dot{E}^{-1/2}R^{-2}\\\label{Appenrho}
&&u=u_{*}\dot{M}^{-1/2}\dot{E}^{1/2}\label{Appenu}.
\end{eqnarray}
Thus these variables can be calculated by
\begin{eqnarray}
&&u_{*}^{2}=2M^{2}/\left(M^{2}+\frac{2}{\gamma-1}\right)\label{dimenless21}\\
&&\rho_{*}=r_{*}/(4\pi u_{*})\label{dimenless22}\\
&&P_{*}=2\rho_{*}/\left[\gamma\left(M^{2}+\frac{2}{\gamma-1}\right)\right]\label{dimenless23}
\end{eqnarray}
for $r<R$ and
\begin{equation}
\rho_{*}=(4\pi r_{*}^{2}u_{*})^{-1}
\end{equation}
but equations (\ref{dimenless21}) and (\ref{dimenless23}) as the same for $r>R$.

Furthermore, if we assume a power-law distributed $q$ and $Q$ for radius-dependent energy-injection and mass-loading efficiencies, i.e., $q\propto r^{-\xi}$ and $Q\propto r^{-\xi}$ with $\xi<3$, the solutions for the Mach number $M$ for $r<R$ becomes 
\begin{equation}
\left[\frac{(\gamma-1)+2/M^{2}}{\gamma+1}\right]^{(1+\gamma)/[2(1+5\gamma-\xi-\xi\gamma)]}
\left[\frac{(3-\xi)\gamma+(1-\xi)/M^{2}}{(1-\xi)+(3-\xi)\gamma}\right]^{-(1+3\gamma-\xi-\xi\gamma)/[(1-\xi)(1+5\gamma-\xi-\xi\gamma)]}=\frac{r}{R},\label{powerlaw1}
\end{equation}
for $\xi\neq 1$, and
\begin{equation}
\left[\frac{(\gamma-1)+2/M^{2}}{\gamma+1}\right]^{(1+\gamma)/8\gamma}\textrm{exp}\left[\frac{1}{4\gamma}\left(1-\frac{1}{M^{2}}\right)\right]=\frac{r}{R},\label{powerlaw2}
\end{equation}
for $\xi=1$. And the dimensionless equation (\ref{dimenless22}) becomes
\begin{equation}
\rho_{*}=\frac{r_{*}^{1-\xi}}{4\pi u_{*}},\label{dimenless24}
\end{equation}
while other solutions are the same as the case of the uniformly distributed $q$ and $Q$. Using these models of $\xi \neq 0$, we have calculated the constraint on $\beta$ for comparison with the results presented in this paper. For example, for the same $R$, SFR, $f_{d}$ and $\alpha=1$, but varying $\xi$ from 0 to 3, we find that constraints on $\beta$ for SFR $\gtrsim1\;M_{\odot}$ yr$^{-1}$ change by less than a factor of 50\%. For a wide range of $\alpha\lesssim2$, the basic conclusion in this paper that $\beta\lesssim1$ for SFR$\gtrsim10\,M_{\odot}$ yr$^{-1}$ in the CC85 model is not changed.

\end{document}